\documentstyle[times,pramana,epsf,floats,amsfonts,amsmath,bm]{ias}
\newtheorem{theorem}{Theorem}
\begin{document}
\mark{{Equations of motion of compact binaries}{Luc Blanchet}}
\title{Equations of motion of compact binaries at the third\\
post-Newtonian order\footnote{To appear in the Proceedings of the
International Conference on Gravitation and Cosmology (ICGC-2004),
edited by B.R. Iyer, V. Kuriakose and C.V. Vishveshwara, published by
Pramana, The Indian Journal of Physics.}}

\author{Luc Blanchet} \address{Institut d'Astrophysique de Paris
(C.N.R.S.),\\ Gravitation et Cosmologie ---
${\mathcal{G}}{\mathbb{R}}\varepsilon{\mathbb{C}}{\mathcal{O}}$,\\
98$^{\text{bis}}$ boulevard Arago, 75014 Paris, France} \pacs{2.0}
\abstract{The equations of motion of two point masses in harmonic
coordinates are derived through the third post-Newtonian (3PN)
approximation. The problem of self-field regularization (necessary for
removing the divergent self-field of point particles) is dealt with in
two separate steps. In a first step the extended Hadamard
regularization is applied, resulting in equations of motion which are
complete at the 3PN order, except for the occurence of one and only
one unknown parameter. In a second step the dimensional regularization
(in $d$ dimensions) is used as a powerful argument for fixing the
value of this parameter, thereby completing the 3-dimensional
Hadamard-regularization result. The complete equations of motion and
associated energy at the 3PN order are given in the case of circular
orbits.}

\maketitle
\section{Introduction}\label{intro}

The third post-Newtonian approximation of general relativity (in short
3PN),\footnote{Following the standard practice \cite{C65}, we refer to
$n$PN as the terms of the order of $1/c^{2n}$ in the equations of
motion, relatively to the Newtonian acceleration.} became famous in
recent years because of its frightening or, depending on one's state
of mind, fascinating intricacy. In particular, the study of this
approximation reached a somewhat paroxysmic stage when it was realized
that the usual self-field regularization, based on Hadamard's concept
of ``\textit{partie finie}'' \cite{Hadamard,Schwartz}, although having
proved to be very efficient up to the 2PN order, fails to provide a
complete answer to the problem at the 3PN order. Indeed it seems to
inexorably yield the appearance of some numerical coefficients which
cannot be determined within the regularization.

Working at such a high approximation level as the 3PN one does not
represent a purely academic exercise. The current network of
laser-interferometric gravitational-wave detectors (notably the
large-scale ones: LIGO and VIRGO) will soon make possible the study of
the inspiral and coalescence of binary systems of neutron stars and
black holes. To extract useful information from the gravitational
waves, theoretical general-relativistic waveforms are used as
templates in these experiments, and it has been demonstrated that
these must be extremely accurate, which means probably as accurate as
the 3PN approximation \cite{3mn,TNaka94,DIS98}. To construct the 3PN
templates one needs to control both the binary's equations of motion,
at the 3PN order relatively to the Newtonian acceleration,
\textit{and} the gravitational radiation field, also consistent at 3PN
order but with respect to the famous Einstein quadrupole formula,
corresponding to the ``Newtonian'' order in the waveform.

In this paper we focus our attention on the problem of motion of a
point mass binary system. The undetermined parameters which appear,
due to Hadamard self-field regularization, are several, but, in fact,
once one has invoked physical arguments to compute some of them, it
remains \textit{one only one} unknown coefficient, the so-called
``static'' ambiguity parameter $\omega_s$ in the 3PN Hamiltonian in
ADM coordinates \cite{JaraS98,JaraS99}, or, equivalently, the
parameter denoted $\lambda$ in the 3PN equations of motion in harmonic
coordinates \cite{BF00,BFeom}. [We mean by physical arguments the
requirement of invariance under global Poincar\'e transformations, and
the demand that the equations of motion should be derivable from a
Lagrangian (neglecting the 2.5PN radiation reaction term).] These
parameters are related to each other by \cite{BF00,DJSequiv,ABF01}
\begin{equation} 
\lambda=-\frac{3}{11}\omega_s-\frac{1987}{3080}\,.
\label{lamboms}
\end{equation}
On the other hand, concerning the radiation field, three other
parameters, $\xi$, $\kappa$ and $\zeta$, coming from the Hadamard
regularization of the 3PN quadrupole moment, appear
\cite{BIJ02}. There is, however, a single parameter which enters the
orbital phase of inspiralling compact binaries, in the form of a
linear combination of $\theta\equiv\xi+2\kappa+\zeta$ and
$\lambda$. [Notice that $\lambda$ enters the radiation field because
of time differentiations of the 3PN quadrupole moment and replacement
of the accelerations by the 3PN equations of motion.]

The regularization ``ambiguities'', say $\omega_s$ or $\lambda$, are
not real physical ambiguities, which would arise, for instance, from
some fundamental failure of the post-Newtonian expansion to
approximate the physics of black holes at high order. Simply, they
reflect some inconsistency, of mathematical origin, in the Hadamard
regularization scheme, when it is applied to the computation of
certain integrals at the 3PN order. Alternatively, one can say that
this regularization, when ``literally'' pushed to its maximum (in the
way proposed in \cite{BFreg,BFregM}), reveals some ``incompleteness''
in making physical predictions, which can or cannot be removed by
external physical arguments. Fortunately, we shall see that the
ambiguity constant (\ref{lamboms}) can be resolved once one disposes
of the appropriate mathematical tools for performing the
regularization.

An improved version of the Hadamard regularization, defined in
\cite{BFreg,BFregM}, is based on: (i) Systematic use of
``partie-finie'' pseudo-functions to represent the functions in the
problem which are singular at the location of the particles; (ii)
Specific distributional derivatives generalizing those of the standard
distribution theory \cite{Schwartz} in order to differentiate the
latter pseudo-functions; (iii) ``Lorentzian'' way of performing the
regularization, defined by the Hadamard partie finie calculated within
the Lorentzian rest frame of the particles. We shall refer to that
regularization \cite{BFreg,BFregM} as the ``extended'' Hadamard (EH)
one.

The EH regularization constitutes the first step of a complete
calculation of the 3PN equations of motion \cite{BFreg,BFregM}. The
second step, aimed at removing the incompleteness $\lambda$, consists
of going to $d$-dimensional space and using complex analytic
continuation in $d$, in what is known as the \textit{dimensional
regularization} (henceforth abbreviated as
``dimreg'').\footnote{Dimreg was invented as a mean to preserve the
gauge symmetry of perturbative quantum gauge field theories
\cite{tHooft}.}  For the moment it is not possible to derive the 3PN
equations of motion in any $d$ dimensions, \textit{i.e.} not
necessarily of the form $d=3+\varepsilon$, where
$\varepsilon\rightarrow 0$. This is why one still has to rely on the
3-dimensional calculation of the equations of motion by means of EH
regularization. This second step (dimensional continuation in $d$) has
already been achieved in the context of the 3PN Hamiltonian in ADM
coordinates, with result \cite{DJSdim}

\begin{equation}
\omega_s = 0\,.
\label{oms0}
\end{equation}
In the present contribution we describe our own application of dimreg
(so to say ``on the top'' of Hadamard's regularization) to the
derivation of the 3PN equations of motion, in the framework of
harmonic coordinates, based on recent work in collaboration with
Damour and Esposito-Far\`ese \cite{BDE04}.

\section{Hadamard regularization of Poisson-like integrals}\label{had}

Let us start by giving some reminders of the way we compute the
Hadamard regularization of some potentials having the form of Poisson
or Poisson-like integrals. Let $F({\mathbf{x}})$ be a smooth function
on ${\mathbb{R}}^3$, except at the value of two singular points
${\mathbf{y}}_1$ and ${\mathbf{y}}_2$, around which it admits some Laurent
expansions of the type ($\forall N\in\mathbb{N}$)
\begin{equation}
F({\mathbf{x}})=\sum_{p_0\leq p\leq N}r_1^p
\mathop{f}_1{}_p({\mathbf{n}}_1)+o(r_1^N)\,,
\label{Fx}
\end{equation}
where $r_1\equiv \vert{\mathbf{x}}-{\mathbf{y}}_1\vert\rightarrow 0$, and
the ${}_1f_p({\mathbf{n}}_1)$'s denote the coefficients of the various
powers of $r_1$, which are functions of the positions and velocities
of the particles, and of the unit direction ${\mathbf{n}}_1\equiv
({\mathbf{x}}-{\mathbf{y}}_1)/r_1$ of approach to singularity 1 (we have
also the same expansion corresponding to the singularity 2). The
powers of $r_1$ are relative integers, $p\in \mathbb{Z}$, bounded from
below by some typically negative $p_0$ depending on the function $F$.

We shall discuss the prescription (taken in \cite{BFreg}) to define
the ``value at ${\mathbf{x}}' = {\mathbf{y}}_1$'' of the singular
Poisson integral $P({\mathbf{x}}')$ of the source function
$F({\mathbf{x}})$. The potential $P({\mathbf{x}}')$ is defined, at any
field point ${\mathbf{x}}'$ different from the singularities, in the
sense of the Hadamard partie-finie (Pf) of an integral, \textit{i.e.}
\begin{equation}
P({\mathbf{x}}')=-\frac{1}{4\pi}{\mathrm{Pf}}_{s_1,s_2}
\int\frac{d^3{\mathbf{x}}}{\vert{\mathbf{x}}-
{\mathbf{x}}'\vert}F({\mathbf{x}})\,.
\label{Px}
\end{equation}
This ``partie finie'' involves two constants, $s_1$ and $s_2$, which
parametrize some logarithmic terms, and are associated with the
characteristics of the regularizing volumes around the two particles,
which have been excised from ${\mathbb{R}}^3$ in order to define the
partie finie by means of the limit, when the size of these volumes
tends to zero, of the integral external to the volumes.

The value at ${\mathbf{x}}' = {\mathbf{y}}_1$ of the function
$P({\mathbf{x}}')$ is defined by the Hadamard partie finie in the
singular limit ${\mathbf{x}}' \rightarrow {\mathbf{y}}_1$, given as
usual by the angular average of the coefficient of the zeroth power of
$r'_1\equiv \vert{\mathbf{x}}'-{\mathbf{y}}_1\vert$ when
$r'_1\rightarrow 0$. Notice first that $P({\mathbf{x}}')$ does not
admit an expansion when $r_1'\rightarrow 0$ of the same type as in
Eq.~(\ref{Fx}), since it involves also a term proportional to the
\textit{logarithm} of $r_1'$. Thus we shall have, rather than a
power-like expansion,
\begin{equation}
P({\mathbf{x}}')=\sum_{p_0'\leq p\leq N}{r_1'}^p
\left[\mathop{g}_1{}_p({\mathbf{n}}_1')+
\mathop{h}_1{}_p({\mathbf{n}}_1')\ln r_1'\right]+o({r_1'}^N)\,,
\label{ExpandPx}
\end{equation}
where the coefficients ${}_1g_p$ and ${}_1h_p$ depend on the angles
${\mathbf{n}}_1'$, and also on the constants $s_1$ and $s_2$, in such a
way that when combining together the terms in (\ref{ExpandPx}) the
constant $r_1'$ always appears in ``adimensionalized'' form like in
$\ln (r_1'/s_1)$. Then we define the Hadamard partie finie at point 1
in the standard way (taking the spherical average of the zeroth-order
power of $r_1'$), except that we include the contribution linked to
the logarithm of $r_1'$, which is possibly present into that
coefficient. More precisely, we define
\begin{equation}
(P)_1 \equiv \bigl<\mathop{g}_1{}_0\bigr>+
\bigl<\mathop{h}_1{}_0\bigr>\ln r_1'\,,
\label{P1def}
\end{equation}
where the brackets denote the angular average, over the solid angle
element $d\Omega ({\mathbf{n}}_1')$ on the unit sphere. Let us
emphasize that in (\ref{P1def}) we have introduced in fact a
\textit{new regularization scale} denoted $r'_1$, which can be seen as
some ``small'' but finite cut-off length scale [so that $\ln r'_1$ in
Eq.~(\ref{P1def}) is a finite, but ``large'' cut-off dependent
contribution]. To compute the partie finie one must apply the
definition (\ref{P1def}) to the Poisson integral (\ref{Px}), which
involves evaluating correctly the angular integration therein. The
result, proved in Theorem 3 of \cite{BFreg}, reads
\begin{equation}
(P)_1=-\frac{1}{4\pi}{\mathrm{Pf}}_{s_1,s_2}
\int\frac{d^3{\mathbf{x}}}{r_1}F({\mathbf{x}})+\left[\ln\left(
\frac{r_1'}{s_1}\right)-1\right]\bigl<\mathop{f}_1{}_{-2}\bigr>\,.
\label{P1}
\end{equation}
The first term is simply the value of the potential at the point 1,
namely $P({\mathbf{y}}_1)$, which would in fact constitute a
``na\"{\i}ve'' way to implement the regularization, but would not
yield 3PN equations of motion compatible with basic physical
properties such as energy conservation. The supplementary term makes
the partie finie to differ from the na\"{\i}ve guess
$P({\mathbf{y}}_1)$ in a way which was found to play a significant
role in the computations of \cite{BFeom}. The apparent dependence of
the result (\ref{P1}) on the scale $s_1$ is illusory. The
$s_1$-dependence of the R.H.S. of Eq.~(\ref{P1}) cancels between the
first and the second terms, so the result depends only on the
constants $r_1'$ and $s_2$, and we have in fact the following simpler
rewriting of (\ref{P1}),
\begin{equation}
(P)_1 = -\frac{1}{4\pi}{\mathrm{Pf}}_{r_1',s_2}
\int\frac{d^3{\mathbf{x}}}{r_1}
F({\mathbf{x}})-\bigl<\mathop{f}_1{}_{-2}\bigr>\, .
\label{P1'}
\end{equation}
Similarly the regularization performed at point 2 will depend on
$r_2'$ and $s_1$, so that the binary's point-particle dynamics depends
on four (\textit{a priori} independent) length scales $r_1'$, $s_2$
and $r_2'$, $s_1$. Because we work at the level of the equations of
motion (instead of, say, the Lagrangian), many of the terms we shall
need are in the form of the \textit{gradient} of a Poisson
potential. For the gradient we have a formula analogous to (\ref{P1'})
and given by Eq.~(5.17a) of \cite{BFreg}, namely
\begin{eqnarray}
(\partial_iP)_1&=&-\frac{1}{4\pi} {\mathrm{Pf}}_{s_1,s_2}\int d^3
{\mathbf{x}}\frac{n_1^i}{r_1^2}F({\mathbf{x}})+\ln\left(
\frac{r_1'}{s_1}\right)\bigl<n_1^i\mathop{f}_1{}_{-1}\bigr>\\
&=&-\frac{1}{4\pi} {\mathrm{Pf}}_{r_1',s_2}\int d^3
{\mathbf{x}}\frac{n_1^i}{r_1^2}F({\mathbf{x}}) \,,
\label{diP1}
\end{eqnarray}
where we have taken into account (in the rewriting of the second line)
the always correct fact that the constant $s_1$ cancels out and gets
``replaced'' by $r_1'$.

We must also treat the more general case of potentials in the form of
retarded integrals, but because we shall have to consider (in Section
\ref{diff} below) only the \textit{difference} between dimreg and the
Hadamard regularization, it will turn out that the first-order
retardation (1PN relative order) is sufficient for this purpose. Here
we are not interested in radiation-reaction effects, so we shall use
the symmetric (half-retarded plus half-advanced) integral. At the 1PN
order we thus have to evaluate
\begin{equation}
R({\mathbf{x}}')=P({\mathbf{x}}')+
\frac{1}{2c^2}Q({\mathbf{x}}')+{\mathcal{O}}\left(\frac{1}{c^4}\right)\,,
\label{RPQ}
\end{equation}
where $P({\mathbf{x}}')$ is given by (\ref{Px}), and where
$Q({\mathbf{x}}')$ denotes (two times) the double or ``twice-iterated''
Poisson integral of the second-time derivative, still endowed with a
prescription of taking the Hadamard partie finie, namely
\begin{equation}
Q({\mathbf{x}}')=-\frac{1}{4\pi}{\mathrm{Pf}}_{s_1,s_2}\int
d^3{\mathbf{x}}\,\vert{\mathbf{x}}-{\mathbf{x}}'
\vert\partial_t^2F({\mathbf{x}})\,.
\label{Qx'}
\end{equation}
In the case of $Q({\mathbf{x}}')$ the results concerning the partie
finie at point 1 were given by Eqs.~(5.16) and (5.17b) of
\cite{BFreg},
\begin{eqnarray}
(Q)_1&=&-\frac{1}{4\pi}{\mathrm{Pf}}_{r_1',s_2}\int d^3{\mathbf{x}}\,r_1
\partial_t^2F({\mathbf{x}})+\frac{1}{2}\bigl<\mathop{k}_1{}_{-4}\bigr>
\,,\\ (\partial_iQ)_1&=&\frac{1}{4\pi}{\mathrm{Pf}}_{r_1',s_2}\int
d^3{\mathbf{x}}\,n_1^i \partial_t^2F({\mathbf{x}})+\frac{1}{2}\bigl<
n_1^i\mathop{k}_1{}_{-3}\bigr>\,,
\label{Q1}\end{eqnarray}
where the ${}_1k_p$'s denote the analogues of the coefficients
${}_1f_p$, parametrizing the expansion of $F$ when $r_1'\rightarrow
0$, but corresponding to the double time-derivative $\partial_t^2F$
instead of $F$. 

There is an important point concerning the treatment of the repeated
time derivative $\partial_t^2F({\mathbf{x}})$ in Eqs.~(\ref{Q1}). As
we are talking here about Hadamard-regularized integrals (which excise
small balls around both ${\mathbf{y}}_1$ and ${\mathbf{y}}_2$), the
value of $\partial_t^2 F({\mathbf{x}})$ can be simply taken in the
sense of ordinary functions, \textit{i.e.}, without including eventual
``distributional'' contributions proportional to $\delta ({\mathbf{x}}
- {\mathbf{y}}_1)$ or $\delta ({\mathbf{x}} - {\mathbf{y}}_2)$ and
their derivatives. However, we know that such terms are necessary for
the consistency of the calculation (without them, for instance, the
calculation would be incorrect already at the 2PN order). In EH
regularization, there is a specific prescription for the
distributional derivative which is issued from the generalized
framework of \cite{BFreg}. In dimreg we shall use simply the standard
Schwartz distributional derivatives in $d$ dimensions. [As it turns
out, the Schwartz derivatives yield some ill-defined (formally
infinite) expressions in 3 dimensions --- this is why a generalization
of the Schwartz distributional derivative defined in \cite{BFreg} was
found to be necessary --- but the latter expressions are proved to be
rigorously zero when computed in $d$ dimensions.] Therefore, in our
computation of the \textit{difference} between dimreg and Hadamard
regularization (next Section), we must also include the difference
between the different prescriptions for the distributional derivatives
in $d$ and in 3 dimensions. We refer to \cite{BDE04} for the details.

\section{Difference between the dimensional and Hadamard regularizations}\label{diff}

In dimreg the computation of the regularized value of Poisson or
Poisson-like integrals is very simple \cite{BDE04}. First of all, the
generalization of the function $F$ in $d$ dimensions will be some
$F^{(d)}$ which admits when $r_1\rightarrow 0$ a more complicated
expansion, namely ($\forall N\in\mathbb{N}$)
\begin{equation}
F^{(d)}({\mathbf{x}})=\sum_{\substack{p_0\leq p\leq N\\ q_0\leq q\leq
q_1}}r_1^{p+q\varepsilon}\mathop{f}_1{}_{p,q}^{
(\varepsilon)}({\mathbf{n}}_1)+o(r_1^N)\,,
\label{Fd}
\end{equation}
where $p$ and $q$ are relative integers ($p,q\in\mathbb{Z}$), whose
values are limited by some $p_0$, $q_0$ and $q_1$ as indicated. The
expansion (\ref{Fd}) differs from the corresponding expansion in 3
dimensions, as given in Eq.~(\ref{Fx}), by the appearance of integer
powers of $r_1^\varepsilon$ where we denote $\varepsilon \equiv
d-3$. The coefficients ${}_1f_{p,q}^{(\varepsilon)}$ depend on the
unit vector ${\mathbf{n}}_1$ in $d$ dimensions, on the positions and
coordinate velocities of the particles, and also on the characteristic
length scale $\ell_0$ of dimensional regularization. The latter can be
introduced into the formalism by saying that the constant $G$ in the
R.H.S. of the Einstein field equations is related to $G_N$, the usual
Newton constant in 3 dimensions, by $G=G_N l_0^\varepsilon$. Because
$F^{(d)}\rightarrow F$ when $d\rightarrow 3$ we necessarily have some
constraint on the coefficients ${}_1f_{p,q}^{(\varepsilon)}$ so that
we are in agreement with the expansion (\ref{Fx}) in this limit.

Consider now the Poisson integral of $F^{(d)}$, in $d$ dimensions,
given by the standard Green's function for the Laplace operator in $d$
dimensions, namely
\begin{equation}
P^{(d)}({\mathbf{x}}')=\Delta^{-1}\bigl[F^{(d)}({\mathbf{x}})\bigr]\equiv
-\frac{\tilde{k}}{4\pi}
\int\frac{d^d{\mathbf{x}}}{\vert{\mathbf{x}}-{\mathbf{x}}'\vert^{d-2}}
F^{(d)}({\mathbf{x}})\,,
\label{Pdx}
\end{equation}
where $\tilde{k}$ is related to the usual Eulerian $\Gamma$-function
by\,\footnote{We have $\lim_{d\rightarrow 3}\tilde{k}=1$. Notice the
following connection to the volume of the sphere with $d-1$ dimensions
(\textit{i.e.}, embedded into Euclidean $d$-dimensional space):
$$\tilde{k}\,\Omega_{d-1}=\frac{4\pi}{d-2}\,.
$$}
\begin{equation}
\tilde{k}=\frac{\Gamma\left(\frac{d-2}{2}\right)}{\pi^{
\frac{d-2}{2}}}\,.
\end{equation}
To evaluate the Poisson integral at the singular point ${\mathbf{x}}' =
{\mathbf{y}}_1$ is quite easy to do in dimreg, because the nice
properties of analytic continuation allow simply to get $[P^{(d)}
({\mathbf{x}}')]_{\mathbf{x}' = {\mathbf{y}}_1}$ by replacing
${\mathbf{x}}'$ by ${\mathbf{y}}_1$ into the explicit integral form
(\ref{Pdx}). So, we simply have,
\begin{equation}
P^{(d)}({\mathbf{y}}_1)=-\frac{\tilde{k}}{4\pi}
\int\frac{d^d{\mathbf{x}}}{r_1^{d-2}}F^{(d)}({\mathbf{x}})\,.\label{Pd}
\end{equation}
Similarly, for the twice iterated Poisson integral, and the relevant
gradients of potentials,
\begin{eqnarray}
Q^{(d)}({\mathbf{y}}_1)&=&-\frac{\tilde{k}}{4\pi(4-d)} \int
d^d{\mathbf{x}}\,r_1^{4-d}\partial_t^2F^{(d)}({\mathbf{x}})\,,\label{Qd}\\
\partial_iP^{(d)}({\mathbf{y}}_1)&=&-\frac{\tilde{k}(d-2)}{4\pi} \int
d^d{\mathbf{x}}\,\frac{n_1^i}{r_1^{d-1}}F^{(d)}({\mathbf{x}})\,,
\label{dPd}\\
\partial_iQ^{(d)}({\mathbf{y}}_1)&=&\frac{\tilde{k}}{4\pi} \int
d^d{\mathbf{x}}\,n_1^i r_1^{3-d}\partial_t^2F^{(d)}({\mathbf{x}})\,.
\label{dQd}
\end{eqnarray}

The main technical step of our strategy will then consist of computing
the \textit{difference} between the $d$-dimensional Poisson-type
potentials (\ref{Pd})--(\ref{dQd}), and their ``pure
Hadamard-Schwartz'' 3-dimensional counterparts, given by expressions
such as (\ref{P1'}). By pure Hadamard-Schwartz (pHS) we mean in some
sense the ``core'' of the Hadamard regularization, \textit{i.e.}
merely based on the usual notion of the partie finie of a singular
function or a divergent integral, but without the improvements brought
about by the EH regularization (see \cite{BDE04} for more
details). For instance, the computations of Section \ref{had} above
belong to the pHS regularization, but the special treatment of
distributional derivatives in three dimensions is specific to the EH
regularization. Given the results $(P)_1$ and
$P^{(d)}({\mathbf{y}}_1)$ of the two regularizations, denoting the
difference by means of the script letter ${\mathcal{D}}$, we thus pose
\begin{equation}
{\mathcal{D}}P(1)\equiv P^{(d)}({\mathbf{y}}_1)-(P)_1\,.
\label{DP1}
\end{equation}
That is, ${\mathcal{D}}P(1)$ is what we shall have to \textit{add} to
the pHS result in order to get the correct $d$-dimensional
result. Note that, in this paper, we shall only compute the first two
terms, $a_{-1} \, \varepsilon^{-1} + a_0 + {\mathcal{O}}
(\varepsilon)$, of the Laurent expansion of ${\mathcal{D}}P(1)$ when
$\varepsilon \rightarrow 0$. [We leave to future work an eventual
computation of the $d$-dimensional equations of motion as an exact
function of the complex number $d$.] This is the information we shall
need to fix the value of the parameter $\lambda$. As we shall see, the
difference ${\mathcal{D}}P(1)$ comes exclusively from the contribution
of poles $\propto 1/\varepsilon$ (and their associated finite part) in
the $d$-dimensional calculation. Here we simply state the result
without proof (see \cite{BDE04} for details). We obtain the following
closed-form expression for the difference [valid up to the neglect of
higher-order terms ${\mathcal{O}}(\varepsilon)$],
\begin{eqnarray}\label{DP1total}
{\mathcal{D}}P(1)=&-&\frac{1}{\varepsilon
(1+\varepsilon)}\!\!\sum_{q_0\leq q\leq q_1}\!\!\left(\frac{1}{q}+\varepsilon
\bigl[\ln r_1'-1\bigr]\right)\bigl<\mathop
{f}_1{}_{-2,q}^{(\varepsilon)}\bigr>\\
&-&\frac{1}{\varepsilon (1+\varepsilon)}\!\!\sum_{q_0\leq q\leq
q_1}\!\!\left(\frac{1}{q+1}+\varepsilon\ln s_2\right)
\sum_{\ell=0}^{+\infty}\frac{(-)^\ell}{\ell!}\partial_L
\left(\frac{1}{r_{12}^{1+\varepsilon}}\right)\bigl<
n_2^L\mathop{f}_2{}_{-\ell-3,q}^{(\varepsilon)}\bigr>\,,\nonumber
\end{eqnarray}
which constitutes the basis of all the practical calculations in the
work \cite{BDE04}.\footnote{With the same notation as in \cite{BDE04}
the multipole expansion in $d$ dimensions reads as
$$r_1^{2-d}=\sum_{\ell=0}^{+\infty}\frac{(-)^\ell}{\ell!}\partial_L
\left(\frac{1}{r_{12}^{1+\varepsilon}}\right)r_2^\ell n_2^L\ .
\label{r1ofr2}
$$} Here we still use the bracket notation to denote the angular
average, but now performed in $d$ dimensions, \textit{i.e.}  over the
solid-angle element $d\Omega_{d-1}$ associated with the
($d-1$)-dimensional sphere. Notice that (\ref{DP1total}) depends on
the two ``constants'' $\ln r_1'$ and $\ln s_2$. As we shall check,
these $\ln r_1'$ and $\ln s_2$ will exactly cancel out the same
constants present in the pHS calculation, so that the dimreg
acceleration will be finally free of the constants $r_1'$ and
$s_2$. Note also that the coefficients ${}_1f_{p,q}^{(\varepsilon)}$
and ${}_2f_{p,q}^{(\varepsilon)}$ in $d$ dimensions depend on the
length scale $\ell_0$ associated with dimreg. Taking this dependence
into account one can verify that $r_1'$ and $s_2$ in (\ref{DP1total})
appear only in the combinations $\ln (r_1'/\ell_0)$ and $\ln
(s_2/\ell_0)$.

Let us give also the formula for the difference between the
\textit{gradients} of potentials, \textit{i.e.}
\begin{equation}
{\mathcal{D}}\partial_iP(1)\equiv\partial_iP^{(d)}({\mathbf{y}}_1)-
(\partial_iP)_1\,,
\label{DdiP1}
\end{equation}
which is readily obtained by the same method. We have
\begin{eqnarray}\label{DdiP1total}
{\mathcal{D}}\partial_iP(1)=&-&\frac{1}{\varepsilon}
\!\!\!\sum_{q_0\leq q\leq q_1}\!\!\!\left(\frac{1}{q}+\varepsilon\ln
r_1'\right)\bigl<n_1^i\mathop{f}_1{}_{-1,q }^{(\varepsilon)}\bigr>\\
&-&\frac{1}{\varepsilon (1+\varepsilon)}\!\!\!\sum_{q_0\leq q\leq
q_1}\!\!\!\left(\frac{1}{q+1}+\varepsilon\ln s_2\right)
\sum_{\ell=0}^{+\infty}\frac{(-)^\ell}{\ell!}\partial_{iL}
\left(\frac{1}{r_{12}^{1+\varepsilon}}\right)\bigl<
n_2^L\mathop{f}_2{}_{-\ell-3,q}^{(\varepsilon)}\bigr>\,.\nonumber
\end{eqnarray}
Formulae (\ref{DP1total}) and (\ref{DdiP1total}) correspond to the
difference of Poisson integrals. But we have already discussed that we
need also the difference of inverse d'Alembertian integrals at the 1PN
order. To express as simply as possible the 1PN-accurate
generalizations of Eqs.~(\ref{DP1total}) and (\ref{DdiP1total}), let
us define two \textit{functionals} ${\mathcal{H}}$ and
${\mathcal{H}}_i$ which are such that their actions on any
$d$-dimensional function $F^{(d)}$ is given by the R.H.S.'s of
Eqs.~(\ref{DP1total}) and (\ref{DdiP1total}), \textit{i.e.}, so that
\begin{eqnarray}
{\mathcal{D}}P(1)&=&{\mathcal{H}}\bigl[F^{(d)}\bigr]\,,
\label{calF}\\
{\mathcal{D}}\partial_iP(1)&=&{\mathcal{H}}_i\bigl[F^{(d)}\bigr]\,.
\label{calFi}
\end{eqnarray}
The difference of 1PN-retarded potentials and gradients of potentials
is denoted
\begin{eqnarray}
{\mathcal{D}}R(1)&\equiv&R^{(d)}({\mathbf{y}}_1)-(R)_1\,,
\label{DR1}\\
{\mathcal{D}}\partial_iR(1)&\equiv&\partial_iR^{(d)}({\mathbf{y}}_1)
-(\partial_iR)_1\,,
\label{DdiR1}
\end{eqnarray}
where in 3 dimensions the potential $R({\mathbf{x}}')$ is defined by
Eq.~(\ref{RPQ}) and the regularized values $(R)_1$ and
$(\partial_iR)_1$ follow from (\ref{P1'}), (\ref{diP1}), (\ref{Q1}),
and where in $d$ dimensions $R^{(d)}({\mathbf{y}}_1)$ and
$\partial_iR^{(d)}({\mathbf{y}}_1)$ are consequences of
Eqs.~(\ref{Pd})--(\ref{dQd}). With this notation we now have our
result, that the difference in the case of such 1PN-expanded
potentials reads in terms of the above defined functionals
${\mathcal{H}}$ and ${\mathcal{H}}_i$ as
\begin{eqnarray}
{\mathcal{D}}R(1)&=&{\mathcal{H}}\left[F^{(d)}+\frac{r_1^2}{2c^2(4-d)}
\partial_t^2F^{(d)}\right]
-\frac{3}{4c^2}\bigl<\mathop{k}_1{}_{-4}\bigr>
+{\mathcal{O}}\left(\frac{1}{c^4}\right),
\label{DR1result}\\
{\mathcal{D}}\partial_iR(1)&=&{\mathcal{H}}_i\left[F^{(d)}
-\frac{r_1^2}{2c^2(d-2)}\partial_t^2F^{(d)}\right]
-\frac{1}{4c^2}\bigl<n_1^i\mathop{k}_1{}_{-3}\bigr>
+{\mathcal{O}}\left(\frac{1}{c^4}\right).
\label{DdiR1result}
\end{eqnarray}
These formulae involve some ``effective'' functions which are to be
inserted into the functional brackets of ${\mathcal{H}}$ and
${\mathcal{H}}_i$. Beware of the fact that the effective functions are
not the same in the cases of a potential and the gradient of that
potential. Note the presence, besides the main terms
${\mathcal{H}}[\cdots]$ and ${\mathcal{H}}_i[\cdots]$, of some extra
terms, purely of order 1PN, in
Eqs.~(\ref{DR1result})--(\ref{DdiR1result}). These terms are made of
the average of some coefficients ${}_1k_p$ of the powers $r_1^p$ in
the expansion when $r_1\rightarrow 0$ of the
\textit{second-time-derivative} of $F$, namely $\partial_t^2F$. They
do not seem to admit a simple interpretation. They are important to
get the final correct result.

\section{Dimensional regularization of the equations of motion}\label{dimreg}

We outline next the way we obtain from the previous computation of the
``difference'' the 3PN equations of motion in dimreg, and show how
they are physically equivalent to the EH-regularized equations of
motion. We start from the end results of \cite{BFeom} for the 3PN
acceleration of the first particle, say ${\mathbf{a}}_1^\text{BF}$,
depending on the two arbitrary length scales $r'_1$ and $r'_2$
(appearing when regularizing Poisson-like integrals in Section
\ref{had}), and on the ``ambiguity'' parameter $\lambda$. Explicitly,
we define
\begin{equation}
{\mathbf{a}}_1^\text{BF} \bigl[\lambda;r_1',r_2'\bigr] \equiv
\text{R.H.S. of Eq.~(7.16) in Ref. \cite{BFeom}}\,.
\label{a1BF}
\end{equation}
Here the acceleration is considered as a function of the two masses
$m_1$ and $m_2$, the relative distance
${\mathbf{y}}_1-{\mathbf{y}}_2\equiv r_{12}{\mathbf{n}}_{12}$ (where
${\mathbf{n}}_{12}$ is the unit vector directed from particle 2 to
particle 1), the two coordinate velocities ${\mathbf{v}}_1$ and
${\mathbf{v}}_2$, and also, as emphasized in (\ref{a1BF}), the parameter
$\lambda$ as well as two regularization length scales $r_1'$ and
$r_2'$. The latter length scales enter the equations of motion at the
3PN level through the logarithms $\ln (r_{12}/r_1')$ and $\ln
(r_{12}/r_2')$. They come from the regularization as the field point
${\mathbf{x}}'$ tends to ${\mathbf{y}}_1$ or ${\mathbf{y}}_2$ of
Poisson-type integrals (see Section \ref{had} above). The length
scales $r_1'$, $r_2'$ are ``pure gauge'' in the sense that they can be
removed by the effect induced on the world-lines of a coordinate
transformation of the bulk metric \cite{BFeom}. 

On the other hand, the dimensionless parameter $\lambda$ entering the
final result (\ref{a1BF}) corresponds to genuine physical effects. It
was introduced by requiring that the 3PN equations of motion admit a
conserved energy (and more generally be derivable from a
Lagrangian). This extra requirement imposed \textit{two relations}
between the two length scales $r'_1$, $r'_2$ and the two other length
scales $s_1$, $s_2$ entering originally into the formalism, namely the
constants $s_1$ and $s_2$ parametrizing the Hadamard partie finie of a
Poisson integral as given by Eq. (\ref{Px}) above. Recall that $s_1$
and $s_2$ are associated with the characteristics of the two
regularizing volumes (notably their shape) around the singularities,
which are excised in order to define the Hadamard partie finie of a
divergent integral. The latter relations were found to be of the form
\begin{equation}
\ln\Bigl(\frac{r_2'}{s_2}\Bigr)=\frac{159}{308}+\lambda
\frac{m_1+m_2}{m_2}
\label{lnr2s2}
\end{equation}
(and $1\leftrightarrow 2$), where the so introduced \textit{single}
dimensionless parameter $\lambda$ has been proved to be a purely
numerical coefficient (\textit{i.e.} independent of the two
masses). It is often convenient to insert Eq. (\ref{lnr2s2}) into
(\ref{a1BF}) and to reexpress the acceleration of particle 1 in terms
of the original regularization length scales entering the Hadamard
regularization of ${\mathbf{a}}_1$, which were in fact $r'_1$ and $s_2$
[as shown, for instance, in Eq.~(\ref{P1'})]. Thus we can consider
alternatively
\begin{equation}
{\mathbf{a}}_1^\text{BF} [r'_1 , s_2] \equiv {\mathbf{a}}_1^\text{BF}
\bigl[\lambda ; r_1', r'_2 (s_2 , \lambda)\bigr]\,,
\label{Foncta1BF}
\end{equation}
where the regularization constants are subject to the constraints
(\ref{lnr2s2}) [we can check that the $\lambda$-dependence on the
R.H.S. of (\ref{Foncta1BF}) disappears when using Eq.~(\ref{lnr2s2})
to replace $r'_2$ as a function of $s_2$ and $\lambda$].

The strategy followed in \cite{BDE04} consists of \textit{two
steps}. The \textit{first step} consists of subtracting all the extra
contributions to Eq.~(\ref{a1BF}), or equivalently
Eq.~(\ref{Foncta1BF}), which were specific consequences of the EH
regularization defined in \cite{BFreg,BFregM}. As has been detailed in
\cite{BDE04}, there are \textit{seven} such extra contributions
$\delta^A {\mathbf{a}}_1$, $A = 1, \cdots , 7$. Subtracting these
contributions boils down to estimating the value of ${\mathbf{a}}_1$
that would be obtained by using a ``pure'' Hadamard regularization,
together with Schwartz distributional derivatives, which is what we
have already called the ``pure Hadamard-Schwartz'' (pHS)
regularization. Such a pHS acceleration was in fact essentially the
result of the first stage of the calculation of ${\mathbf{a}}_1$, as
reported in the (unpublished) thesis \cite{FayeThesis}. It is given by
\begin{equation}
{\mathbf{a}}_1^\text{pHS}\bigl[r_1',s_2\bigr]={\mathbf{a}}_1^\text{BF}
[r'_1 , s_2]-\sum_{A=1}^7\delta^A{\mathbf{a}}_1\,.
\label{accpH}
\end{equation}
The \textit{second step} of our method consists of evaluating the
Laurent expansion, in powers of $\varepsilon = d-3$, of the difference
between the dimreg and pHS (3-dimensional) computations of the
acceleration ${\mathbf{a}}_1$. As we have seen in Section \ref{diff}
this difference makes a contribution only when a term generates a
\textit{pole} $\sim 1/\varepsilon$, in which case dimreg adds an extra
contribution, made of the pole and the finite part associated with the
pole [we consistently neglect all terms
${\mathcal{O}}(\varepsilon)$]. One must then be especially wary of
combinations of terms whose pole parts finally cancel (``cancelled
poles'') but whose dimensionally regularized finite parts generally do
not, and must be evaluated with care. We denote the above defined
difference
\begin{equation}
{\mathcal{D}}{\mathbf{a}}_1={\mathcal{D}}{\mathbf{a}}_1
\bigl[\varepsilon,\ell_0;r_1',s_2\bigr]\equiv{\mathcal{D}}{\mathbf{a}}_1
\bigl[\varepsilon,\ell_0;\lambda ; r_1', r'_2\bigr]\,.
\label{deltaacc}
\end{equation}
It depends both on the Hadamard regularization scales $r_1'$ and $s_2$
(or equivalently on $\lambda$ and $r_1'$, $r_2'$) and on the
regularizing parameters of dimreg, namely $\varepsilon$ and the
characteristic length $\ell_0$. It is made of the sum of all the
individual contributions of the Poisson or Poisson-like integrals as
computed in Section \ref{diff} above [\textit{e.g.}
Eqs. (\ref{DP1total}) and (\ref{DdiP1total})]. Finally, our main
result will be the explicit computation of the $\varepsilon$-expansion
of the dimreg acceleration as
\begin{equation}
{\mathbf{a}}_1^\text{dimreg} [\varepsilon , \ell_0] =
{\mathbf{a}}_1^\text{pHS} [r'_1 , s_2] + {\mathcal{D}}{\mathbf{a}}_1
[\varepsilon , \ell_0 ; r'_1 , s_2]\,.
\label{a1DimReg}
\end{equation}
With this result in hands, we have proved \cite{BDE04} two theorems.

\begin{theorem}
The pole part $\propto 1/\varepsilon$ of the dimreg acceleration
(\ref{a1DimReg}), as well as of the metric field $g_{\mu\nu}(x)$
outside the particles, can be re-absorbed (\textit{i.e.}, renormalized
away) into some shifts of the two ``bare'' world-lines: ${\mathbf{y}}_a
\rightarrow {\mathbf{y}}_a+{\bm{\xi}}_a$, with, say, ${\bm{\xi}}_a \propto
1/\varepsilon$ (``minimal subtraction''; MS), so that the result,
expressed in terms of the ``dressed'' quantities, is finite when
$\varepsilon\rightarrow 0$.
\label{th1}
\end{theorem}
The situation in harmonic coordinates is to be contrasted with the
calculation in ADM-type coordinates within the Hamiltonian formalism
\cite{DJSdim}, where it was shown that all pole parts directly cancel
out in the total 3PN Hamiltonian (no shifts of the world-lines were
needed). The central result of the paper is then as follows.

\begin{theorem}
The ``renormalized'' (finite) dimreg acceleration is physically
equivalent to the EH-regularized acceleration (end result of
Ref.~\cite{BFeom}), in the sense that there exist some shift vectors
${\bm{\xi}}_1 (\varepsilon , \ell_0 ; r'_1)$ and $\bm{\xi}_2
(\varepsilon , \ell_0 ; r'_2)$, such that
\begin{equation}
{\mathbf{a}}_1^{\mathrm{BF}} [\lambda, r'_1 , r'_2] =
\lim_{\varepsilon\rightarrow 0} \,
\bigl[{\mathbf{a}}_1^{\mathrm{dimreg}} [\varepsilon , \ell_0] +
\delta_{\bm{\xi} (\varepsilon , \ell_0 ; r'_1 , r'_2)} \,
{\mathbf{a}}_1 \bigr]
\label{eta}
\end{equation}
(where $\delta_{\bm{\xi}} \, {\mathbf{a}}_1$ denotes the effect of the
shifts on the acceleration\,\footnote{When working at the level of the
equations of motion (not considering the metric outside the
world-lines), the effect of shifts can be seen as being induced by a
coordinate transformation of the bulk metric as in
Ref.~\cite{BFeom}.}), if and only if the heretofore unknown
parameter $\lambda$ entering the harmonic-coordinates equations of
motion takes the value
\begin{equation}
\lambda = -\frac{1987}{3080}\,.
\label{lambda}
\end{equation}
\label{th2}
\end{theorem}
The precise shifts $\bm{\xi}_a (\varepsilon)$ needed in Theorem
\ref{th2} involve not only a pole contribution $\propto 1/\varepsilon$,
which defines the ``minimal'' (MS) shifts considered in Theorem
\ref{th1}, but also a finite contribution when $\varepsilon\rightarrow
0$. Their explicit expressions read:
\begin{equation}
\bm{\xi}_1=\frac{11}{3}\frac{G_N^2\,m_1^2}{c^6}\left[
\frac{1}{\varepsilon}-2\ln\left(
\frac{r'_1\overline{q}^{1/2}}{\ell_0}\right)
-\frac{327}{1540}\right] {\mathbf{a}}_{N1}~~\text{and}~~1\leftrightarrow
2\,,
\end{equation}
where $G_N$ is the usual Newton's constant, ${\mathbf{a}}_{N1}$
denotes the acceleration of the particle 1 (in $d$ dimensions) at the
Newtonian level, and $\overline{q}\equiv 4\pi e^C$ depends on the
Euler constant $C=0.577\cdots$.

An alternative way to phrase the result (\ref{eta})--(\ref{lambda}),
is to combine Eqs.~(\ref{accpH}) and (\ref{a1DimReg}) in order to
arrive at
\begin{equation}
\lim_{\varepsilon\rightarrow 0} \, \Bigl[{\mathcal{D}}{\mathbf{a}}_1
\bigl[\varepsilon,\ell_0; \hbox{$-\frac{1987}{3080}$} ; r_1',
r'_2\bigr] + \delta_{\bm{\xi} (\varepsilon , \ell_0 ; r'_1 , r'_2)} \,
{\mathbf{a}}_1 \Bigr] = \sum_{A=1}^7\delta^A{\mathbf{a}}_1\,.
\label{equiveta}
\end{equation}
Under this form one sees that the sum of the additional terms
$\delta^A{\mathbf{a}}_1$ differs by a mere shift, \textit{when and
only when} $\lambda$ takes the value (\ref{lambda}), from the specific
contribution ${\mathcal{D}}{\mathbf{a}}_1$, which comes directly from
dimreg. Therefore one can say that, when $\lambda =
-\frac{1987}{3080}$, the EH regularization \cite{BFreg,BFregM} is in
fact (physically) equivalent to dimreg. However the EH regularization
is incomplete, both because it is \textit{a priori} unable to
determine $\lambda$, and also because it necessitates some
``external'' requirements such as the imposition of the link
(\ref{lnr2s2}) in order to ensure the existence of a conserved energy
--- and in fact of the ten first integrals linked to the Poincar\'e
group. By contrast dimreg succeeds automatically (without extra
inputs) in guaranteeing the existence of the ten conserved integrals
of the Poincar\'e group, as already found in Ref.~\cite{DJSdim}.

In view of the necessary link (\ref{lamboms}) provided by the
equivalence between the ADM-Hamiltonian and the harmonic-coordinates
equations of motion, our result (\ref{lambda}) is in perfect agreement
with the result $\omega_s=0$ obtained in \cite{DJSdim}. [One may
wonder why the value of $\lambda$ is a complicated rational fraction
while $\omega_s$ is so simple. This is because $\omega_s$ was
introduced precisely to measure the amount of ambiguities of certain
integrals, while, by contrast, $\lambda$ has been introduced as the
only possible unknown constant in the link between the four arbitrary
scales $r'_1 , r'_2 , s_1 , s_2$ (which has \textit{a priori} nothing
to do with ambiguities of integrals), in a framework where the use of
the EH regularization makes in fact the calculation to be
unambiguous.] Besides the confirmation of the value of $\omega_s$ or
$\lambda$, this result provides a confirmation of the
\textit{consistency} of dimreg, because our explicit calculations are
entirely different from the ones of \cite{DJSdim}: We use harmonic
coordinates (instead of ADM-type ones), we work at the level of the
equations of motion (instead of the Hamiltonian), we use a different
form of Einstein's field equations and we solve them by a different
iteration scheme. Our result is also in agreement with the recent
finding of Refs. \cite{itoh1,itoh2} (see also \cite{IFA01}), where the
3PN equations of motion are derived in harmonic gauge using a
``surface-integral'' approach, aimed at describing \textit{extended}
relativistic compact binary systems in the strong-field point particle
limit.

\section{Equations of motion of circular-orbit 
compact binaries}\label{circ}

From a practical point of view, the determination of the value of
$\lambda$ allows one to use the full 3PN accuracy in the analytical
computation of the dynamics of the last orbits of binary systems
\cite{DJSisco,B02ico}. We assume a circular orbit since most
inspiralling compact binaries will have been circularized at the time
when they enter the frequency bandwidth of the detectors LIGO and
VIRGO. In the case of circular orbits --- apart from the gradual 2.5PN
radiation-reaction inspiral --- the quite complicated equations of
motion, Eq.~(7.16) in Ref. \cite{BFeom}, simplify drastically.

We translate the origin of coordinates to the binary's center-of-mass
by imposing that the binary's center-of-mass vector, deduced from the
Lagrangian formulation of the 3PN equations of motion, is zero (see
\textit{e.g.} Ref. \cite{BI03}). Then, in the center-of-mass frame,
the relative acceleration ${\mathbf{a}}_{12}\equiv
{\mathbf{a}}_1-{\mathbf{a}}_2$ of two bodies moving on a circular
orbit at the 3PN order is given by

\begin{equation}\label{acc}
{\mathbf{a}}_{12} = -\omega^2 {\mathbf{y}}_{12}-
\frac{32}{5}\frac{G^3m^3\nu}{c^5r_{12}^4}{\mathbf{v}}_{12} + {\cal
O}\left(\frac{1}{c^7}\right)\;,
\end{equation}
where ${\mathbf{y}}_{12}\equiv {\mathbf{y}}_1-{\mathbf{y}}_2$ is the
relative separation (in harmonic coordinates) and $\omega$ denotes the
angular frequency of the circular motion; the second term in
Eq. (\ref{acc}), opposite to the velocity ${\mathbf{v}}_{12}\equiv
{\mathbf{v}}_1-{\mathbf{v}}_2$, is the 2.5PN radiation reaction
force. In (\ref{acc}) we have introduced, in addition to the total
mass $m=m_1+m_2$, the symmetric mass ratio
\begin{equation}\label{nu}
\nu\equiv \frac{m_1m_2}{m^2}\;,
\end{equation}
which is generally very useful because of its interesting range of
variation $0<\nu\leq \frac{1}{4}$, with $\nu=\frac{1}{4}$ in the case
of equal masses, and $\nu\to 0$ in the ``test-mass'' limit for one of
the bodies. The main content of the 3PN equations (\ref{acc}) is the
relation between the frequency $\omega$ and the orbital separation
$r_{12}$, that we find to be given by the 3PN-generalized ``Kepler''
third law \cite{BF00,BFeom}
\begin{eqnarray}\label{om2}
\omega_{\mathrm{3PN}}^2 &=& \frac{Gm}{r_{12}^3} \bigg\{ 1+(-3+\nu)
\gamma + \left( 6 +\frac{41}{4} \nu +\nu^2 \right) \gamma^2\\ &+&
\left( -10 + \left[-\frac{75707}{840}+\frac{41}{64}\pi^2
+22\ln\left(\frac{r_{12}}{r'_0}\right)\right]\nu
+\frac{19}{2}\nu^2+\nu^3 \right) \gamma^3 \biggr\}\;,\nonumber
\end{eqnarray}
in which we employ, in order to display the successive post-Newtonian
corrections, the post-Newtonian parameter [of the order of
${\mathcal{O}}(1/c^2)$]
\begin{equation}\label{gamma}
\gamma = \frac{G m}{r_{12}c^2}\;.
\end{equation}
The acceleration (\ref{acc})--(\ref{gamma}) is entirely specified at
the 3PN order except for some unphysical \textit{gauge freedom},
parametrized by the length scale $r'_0$ appearing in Eq.~(\ref{om2}),
which is nothing but the ``logarithmic barycenter'' of the two
gauge-constants $r'_1$ and $r'_2$ entering the end results of
\cite{BFeom}, \textit{i.e.}
\begin{equation}\label{lnr'}
\ln r'_0 = \frac{m_1}{m} \ln r'_1 + \frac{m_2}{m} \ln r'_2\;.
\end{equation}
As for the binary's energy (in the center-of-mass frame), it is
readily obtained from the circular-orbit reduction of the conserved
energy associated with the 3PN Lagrangian in harmonic coordinates
\cite{ABF01}. We find
\begin{eqnarray}\label{Egam}
E_{\mathrm{3PN}} &=& -\frac{\mu c^2 \gamma}{2} \biggl\{ 1
 +\left(-\frac{7}{4}+\frac{1}{4}\nu\right) \gamma +
 \left(-\frac{7}{8}+\frac{49}{8}\nu +\frac{1}{8}\nu^2\right)
 \gamma^2\\
 &&+\left(-\frac{235}{64}+\left[\frac{46031}{2240}-\frac{123}{64}\pi^2
 +\frac{22}{3}\ln \left(\frac{r_{12}}{r_0'}\right)\right]\nu
 +\frac{27}{32}\nu^2 +\frac{5}{64}\nu^3\right)\gamma^3
 \biggr\}\;.\nonumber
\end{eqnarray}
This expression is that of a physical observable $E$, however it
depends on the choice of a coordinate system, because it involves the
post-Newtonian parameter $\gamma$ defined from the harmonic-coordinate
separation $r_{12}$. But the {\it numerical} value of $E$ should not
depend on the choice of a coordinate system, so $E$ must admit a
frame-invariant expression, the same in all coordinate systems. To
find it we re-express $E$ with the help of a frequency-related
parameter $x$ instead of the separation-related parameter $\gamma$
[this is always a good thing to do]. We define $x$ to be, like for
$\gamma$, of the order of ${\mathcal{O}}(1/c^2)$ by posing
\begin{equation}\label{x}
x = \left(\frac{G m \,\omega}{c^3}\right)^{2/3}\,.
\end{equation}
Then we readily obtain the expression of $\gamma$ in terms of $x$ at
3PN order,
\begin{eqnarray}\label{gam3PN}
\gamma_{\mathrm{3PN}} &=& x \biggl\{1+\left(1-\frac{\nu}{3}\right)x +
\left(1-\frac{65}{12} \nu \right) x^2 \\ &+&
\left(1+\left[-\frac{2203}{2520}-\frac{41}{192}\pi^2-\frac{22}{3}
\ln\left(\frac{r_{12}}{{r'}_0}\right)\right]\nu +\frac{229}{36}\nu^2
+\frac{1}{81}\nu^3\right)x^3 \biggr\}\;,\nonumber
\end{eqnarray} 
that we substitute back into Eq. (\ref{Egam}), making all appropriate
post-Newtonian re-expansions. As a result we gladly discover that the
logarithms together with their associated gauge constant $r'_0$ have
cancelled out. Therefore our final result is
\begin{eqnarray}\label{Ex}
E_{\mathrm{3PN}} = &-&\frac{\mu c^2 x}{2} \biggl\{ 1
 +\left(-\frac{3}{4}-\frac{1}{12}\nu\right) x +
 \left(-\frac{27}{8}+\frac{19}{8}\nu -\frac{1}{24}\nu^2\right) x^2\\
 &+&\left(-\frac{675}{64}+\left[\frac{34445}{576}
 -\frac{205}{96}\pi^2\right]\nu -\frac{155}{96}\nu^2
 -\frac{35}{5184}\nu^3\right)x^3 \biggr\}\;.\nonumber
\end{eqnarray}
In the test-mass limit $\nu\to 0$, we recover the energy of a particle
with mass $\mu=m\nu$ in a Schwarzschild background of mass $m$,
i.e. $E_{\rm test}=\mu c^2\left[(1-2x)(1-3x)^{-1/2}-1\right]$, when
developed to 3PN order. Of course, the subtleties we have discussed,
linked with the self-field regularization, disappear in the test-mass
limit, but, interestingly enough, they affect only the term
proportional to $\nu$ in the 3PN coefficient of Eq. (\ref{Ex}); the
terms proportional to $\nu^2$ and $\nu^3$ in Eq. (\ref{Ex}) have been
found to be ``complete'' in EH regularization.

\begin{figure}[h]
\centerline{\epsfxsize=20pc \epsfbox{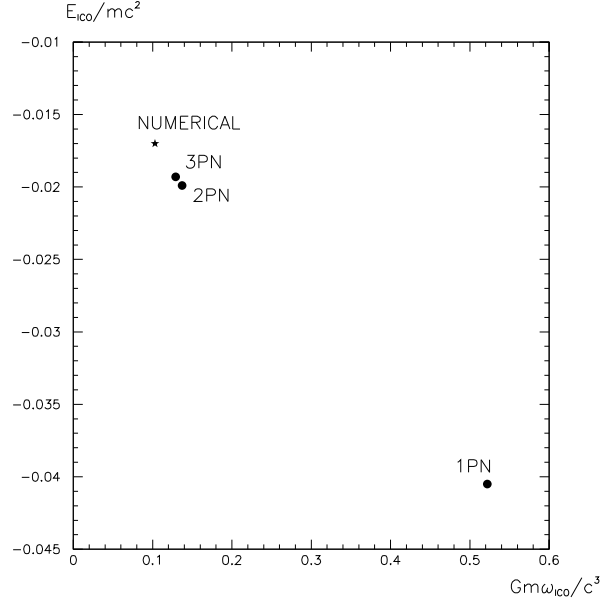}}
\caption{Results for $E_{\rm ICO}$ in terms of $\omega_{\rm ICO}$ for
equal-mass binaries ($\nu=\frac{1}{4}$). The asterisk marks the result
calculated by numerical relativity.
\label{fig}}
\end{figure}

Finally let us compute the innermost circular orbit (ICO) of
point-particle binaries through 3PN order, following
\cite{B02ico}. The ICO is defined as the minimum, when it exists, of
the binary's circular-orbit energy function (\ref{Ex}). In particular,
we do not define the ICO as a point of dynamical
(general-relativistic) unstability. [See Section 6 of \cite{BI03} for
a discussion of the dynamical unstability in the post-Newtonian
framework.] In Fig. \ref{fig} we plot $E_{\mathrm ICO}$ versus
$\omega_{\mathrm ICO}$ in the case of equal masses
($\nu=\frac{1}{4}$), and compare the values with the recent finding of
numerical relativity, obtained by means of a sequence of
quasi-equilibrium configurations under the assumptions of helical
symmetry and conformal flatness \cite{GGB1,GGB2}. As we can see the
2PN and 3PN points are rather close to each other and to the numerical
value. However, the 1PN approximation is clearly not precise enough,
but this is not very surprising in the highly relativistic regime of
the ICO where the orbital velocity reaches $v/c\sim (G
m\,\omega_{\mathrm ICO}/c^3)^{1/3}\sim 0.5$. A striking fact from
Fig. \ref{fig} is that the post-Newtonian series seems to ``converge
well'', but actually the series could be only asymptotic (hence
divergent), and, of course, still give excellent results provided that
the series is truncated near some optimal order of approximation.

Our conclusions, therefore, are that (1) the post-Newtonian
approximation is likely to be valid and quite accurate in the regime
of the ICO (in the equal-mass case), and (2) it is in good agreement
with the result of numerical relativity. Note that the conclusion (1)
contradicts some earlier prejudices about the slow convergence of the
post-Newtonian approximation (see \textit{e.g.}
Ref. \cite{3mn}). Furthermore, our computations are based on the
standard post-Newtonian expansion, without using any resummation
techniques such as Pad\'e approximants and/or effective-one-body
method. For recent comparisons of the post-Newtonian and numerical
calculations in the regime of the ICO, including finite-size effects
appropriate to neutron-star binaries, see \cite{MW02,MW03}.

\acknowledgements

I wish to thank the organizers of an interesting conference. It is a
pleasure to thank my colleagues Thibault Damour and Gilles
Esposito-Far\`ese for the collaboration \cite{BDE04} reported in
Sections \ref{diff} and \ref{dimreg} of this paper.

\end{document}